\documentclass{article}

\newcommand{\p}[1]{(\ref{#1})}

\newcommand{\cD}{{\cal D}}
\newcommand{\cK}{{\cal K}}
\newcommand{\cH}{{\cal H}}
\newcommand{\cU}{{\cal U}}
\newcommand{\cQ}{{\cal Q}}
\newcommand{\cS}{{\cal S}}

\newcommand{\bpsi}{{\bar\psi}{}}

\newcommand{\be}{\begin{equation}}
\newcommand{\ee}{\end{equation}}
\newcommand{\bea}{\begin{eqnarray}}
\newcommand{\eea}{\end{eqnarray}}

\newcommand{\ba}{\begin{array}} \newcommand{\ea}{\end{array}}

\def\im{{\rm i}}

\newcommand{\nn}{\nonumber}

\usepackage{amscd,amsmath,amssymb}

\begin{document}
\thispagestyle{empty}
\vspace{2cm}
\begin{flushright}
%Draft 1 \\
%\today\\
\end{flushright}\vspace{2cm}
\begin{center}
{\Large\bf N=2 supersymmetric particle near extreme Kerr throat}
\end{center}
\vspace{1cm}

\begin{center}
{\Large
Stefano Bellucci$\,{}^{a}$ and Sergey Krivonos$\,{}^{b}$
}\\
\vspace{1.0cm}
${}^a$ {\it
INFN-Laboratori Nazionali di Frascati,
Via E. Fermi 40, 00044 Frascati, Italy} \vspace{0.2cm}

${}^b$
{\it Bogoliubov  Laboratory of Theoretical Physics,
JINR, 141980 Dubna, Russia}
\vspace{0.2cm}
\end{center}
\vspace{2cm}

\begin{abstract}\noindent
We construct a new $N=2$ supersymmetric extension of a massive
particle moving near the horizon of the extreme Kerr black hole.
Our supercharges and Hamiltonian contain the proper number of
fermions (two for each bosonic variables). The key ingredient of
our construction is a proper choice of the bosonic variables which
all have a clear geometric meaning.
\end{abstract}

\newpage
\setcounter{page}{1}
\setcounter{equation}{0}
\section{Introduction}
In the recent paper \cite{AG} the $N=2$ supersymmetric extension
of a massive particle moving near the horizon of the extreme Kerr
black hole has been constructed. The corner stone of this paper is
the new solution of the vacuum Einstein equations found by Bardeen
and Horowitz \cite{BH}. This solution describes the near horizon
geometry of  the extreme Kerr black hole. Very similarly to the
extreme Reissner-Nordstr\"{o}m black hole and its near horizon
limit, the Bertotti-Robinson solution, which has an enhanced
isometry group, the solution constructed in \cite{BH} also has an
enhanced isometry group -- the $SO(1,2)\times U(1)$ one. The
appearance of the $SO(1,2)$ subgroup, which is just a conformal
group in one dimension, is very important -- one may conjecture
that there is some dual conformal field theory description of this
extreme Kerr black hole.

Motivated by the fact that the enhanced isometry group
$SO(1,2)\times U(1)$ of the Bardeen and Horowitz solution
coincides with the bosonic part of the $N=2$ superconformal group
in one dimension, a $N=2$ supersymmetric extension of the massive
particle moving in this background has been derived\footnote{A particle moving in Bertotti-Robinson
space-time admits $N=4$ superconformal symmetry \cite{b1,b2}.}. The explicit
form of the supercharges and the corresponding Hamiltonian
presented in \cite{AG} immediately raise the question -- how can
it be that the number of fermionic components is just two, while
in the theory we have three bosonic fields? This fact is in
contradiction with all known one-dimensional supersymmetric
systems, in which the number of fermionic components is always
larger than (or equal to) the number of bosonic ones.
Nevertheless, the supercharges in \cite{AG} perfectly commute to
span, together with the Hamiltonian and $U(1)$ current, the $N=2$
superconformal group $SU(1,1|1)$.

In the present paper we resolve this paradox and construct a new
variant of the $N=2$ supersymmetric massive particle moving near
the horizon of the extreme Kerr black hole. Roughly speaking, the
situation looks as follows. Firstly, for any one-dimensional
conformal invariant system there is a preferable coordinate system
in which the Hamiltonian reads \cite{AandAll}: \be\label{i1}
H=\frac{P_R^2}{2}+\frac{J}{R^2}, \ee where $R$ is a ``radial''
variable\footnote{It is just equal to $R=e^{\frac{u}{2}}$, where
$u$ is a standard dilaton field}, $P_R$ is the corresponding
momentum, while $J$ describes the ``angular'' part of the system.
The form of the Hamiltonian in \p{i1} is the visualization of the
commonly known fact that any system with the Hamiltonian $J$ can
be made conformal invariant by introducing a proper coupling with
the dilaton. Clearly, the ``angular'' and ``radial'' parts in \p{i1} are
completely independent \be \left\{ P_R, J\right\} = \left\{ R, J
\right\} =0. \ee Secondly, when we turn to derive a supersymmetric
extension of the system in \p{i1} there are two possibilities. The
simplest one is to treat the ``angular'' part of the system $J$ as a
constant which does not participate in the supersymmetrization.
With such approach any inner structure of the ``angular'' part is
completely irrelevant -- the final supercharges and Hamiltonian
will have the same form depending on $J$ as on a coupling
constant. Just this variant of supersymmetrization has been
constructed in \cite{AG}. It is worth to note that the resulting
system is nothing but just a supersymmetric conformal mechanics.
Moreover, one may increase the number of supersymmetries to four,
or even to an arbitrary even number $N=2k$ \cite{leva1,AN1}. The
corresponding superconformal group will be the $SU(1,1|k)$ one.
One should stress that, despite the fact that such an extended
supersymmetric system has a $SO(1,2) \times U(1) \times SU(k)$
bosonic subgroup, there is no contradiction with the isometry
group $SO(1,2) \times U(1)$ of the background metric we started
with: the additional group $SU(k)$ is realized on the fermionic
components only. It is just invisible while we are dealing with
the bosonic sector only. Clearly, it is not possible to establish
any link between supersymmetry charges and Killing spinors within
the proposed supersymmetrization scheme, because the ``angular''
sector left untouched by supersymmetry.

Another variant of the supersymmetrization of the system with the
Hamiltonian \p{i1} includes the supersymmetrization of the ``radial''
as well as ``angular'' sectors of the model. In this case the
restrictions imposed by the existence of the supercharges are
stronger, and the maximal supersymmetry which can be achieved is
indeed the $N=2$ one. The number of physical fermions in the
resulting system will be six - two fermionic components per each
bosonic ones, as it should be in the $N=2$ supersymmetric
one-dimensional model. The construction of such a supersymmetric
system is the subject of the present paper.

The paper is organized as follows. In section 2 we review the
results obtained in \cite{BH,AG}. In addition we also present four
additional multi-parameters solution of the vacuum Einstein
equations which are very similar to those obtained in \cite{BH}.
In section 3 we introduce new variables, in order to bring the
bosonic Hamiltonian to the form \p{i1}. The supercharges and the
Hamiltonian are derived in section 4. We conclude with a short
discussion. \setcounter{equation}{0}
\section{Preliminaries}
The metric we are interesting in, has the form
\cite{BH}\footnote{We fixed one inessential parameter $r_0$
originally present in \cite{BH} to be $r_0^2=2 M^2=1$.}
\be\label{metric1} ds^2=f_1 \left[ -r^2 d\tau^2 +\frac{1}{r^2}
dr^2+d\theta^2 \right]+ f_2\left[ d\phi +r d\tau \right]^2, \ee
where \be\label{sol1} f_1 =\frac{1+ \cos^2 \theta}{2}, \quad
f_2=\frac{ 2 \sin^2 \theta}{1+ \cos^2 \theta}. \ee The metric
\p{metric1} is the solution of the vacuum Einstein equations
\cite{BH}. In principle, forgetting about the former source of the
metric \p{metric1}, one may find some additional metrics still
obeying the vacuum Einstein equations. Indeed, if we let the
functions $f_1,f_2$ in \p{sol1} be arbitrary, depending on the
coordinate $\theta$ only, then, the condition that the metric
\p{metric1} obeys the vacuum Einstein equations results in the
following restrictions on these functions: \bea
&& 4 f_1^2 +3 \left( f_1'\right)^2 +2 f_1 f_1'' -2 \left( f_1 ''\right)^2 +3 f_1' f_1^{(3)} =0, \label{masterEq} \\
&& f_2 = \frac{1}{3 f_1}\left( 4 f_1^2 -3\left( f_1'\right)^2 +4
f_1 f_1'' \right). \label{f2} \eea Among the solutions of the
equation \p{masterEq} there is subset which is similar to
\p{sol1}. This subset of solutions can be selected by the
following Ansatz \be\label{f1} f_1=a_1+a_2 \sin(\theta) +a_3
\cos(\theta)+a_4 \sin(2 \theta) +a_5 \cos(2 \theta) \ee which
yields five solutions\footnote{For each solution with a given set
of parameters one has check that $f_2 \neq 0$. Otherwise, the
metric \p{metric1} will be degenerate.} \be\label{f1f2}
\begin{array}{|l||c|c|c|c|c|}
\hline
sol\phantom{\Biggl(}& a_1&a_2&a_3&a_4&a_5\\
\hline\hline
I\phantom{\Bigl(}&-3 \sqrt{a_4^2+a_5^2}& -\frac{a_3 a_5}{a_4} -\frac{a_3\sqrt{a_4^2+a_5^2}}{a_4}& a_3 & a_4 &a_5 \\
\hline
II\phantom{\Bigl(}&3 \sqrt{a_4^2+a_5^2}& -\frac{a_3 a_5}{a_4} +\frac{a_3\sqrt{a_4^2+a_5^2}}{a_4}& a_3 & a_4 &a_5 \\
\hline
III\phantom{\Bigl(}& 0 & a_2 & a_3 &0 & 0 \\
\hline
IV\phantom{\Bigl(}& 3 a_5 & 0 & a_3 &0 & a_5 \\
\hline
V\phantom{\Bigl(}& -3 a_5 & a_2 & 0 &0 & a_5 \\
\hline
\end{array}
\ee The metric in \p{metric1} corresponds to  solution IV with
$a_5=\frac{1}{4}$ and $a_3=0$.

All these metrics share the same property -- they are invariant
with respect to  $SO(1,2)$ transformations realized in an unusual
way as \cite{BH} \be\label{conf1} \delta \tau = a + b\, \tau +
c\,\left( \tau^2+\frac{1}{r^2}\right), \quad \delta r = - b\, r -
2 c\, \tau\, r,\quad \delta \phi = -\frac{2 c}{r},\qquad \delta
\theta=0, \ee where $a,b,c$ are the parameters of $SO(1,2)$.

Just the presence of this symmetry was the main motivation in the
paper \cite{AG} where the static gauge action of a relativistic
particle moving in the background with the metric \p{metric1} has
been written as \be\label{sAG} S=- m \int  \sqrt{-ds^2}=-m \int
d\tau \sqrt{f_1\left( r^2 -\frac{{\dot
r}{}^2}{r^2}-{\dot\theta}{}^2\right) -f_2 \left( \dot\phi
+r\right)^2}. \ee With this action, the Hamiltonian, together with
the first integrals generating the conformal transformations
\p{conf1}, were constructed as \bea\label{genAG}
H&=& r \left( \sqrt{ m^2 f_1 +\left( r p_r\right)^2 +p_\theta^2 +\frac{f_1}{f_2} p_\phi^2}-p_\phi \right), \nn\\
K&=&\frac{1}{r} \left( \sqrt{ m^2 f_1 +\left( r p_r\right)^2 +p_\theta^2 +\frac{f_1}{f_2} p_\phi^2}+p_\phi\right) +\tau^2 H+ 2 \tau r p_r, \nn\\
D&=& r p_r +\tau H.
\eea
Under the Poisson brackets
\be\label{pbAG}
\left\{r ,p_r\right\}=1,\quad \left\{\phi ,p_\phi\right\}=1,\quad \left\{\theta ,p_\theta\right\}=1,
\ee
they form $so(1,2)$ algebra
\be\label{sl2AG}
\left\{ H,D\right\}=H, \quad \left\{ H,K\right\}=2 D, \quad\left\{ D,K\right\}=K.
\ee
One additional integral of motion fully commuting with $(H,D,K)$, is $p_\phi$.

In the same paper \cite{AG} the $N=2$ supersymmetric extension of
the action \p{sAG} has been constructed. In the subsequent
sections we will construct the extended version of the
supersymmetric mechanics of \cite{AG}. However, before passing to
the supersymmetrization procedure it makes sense to change our
coordinates (fields) to bring the $SO(1,2)$ transformations to a
more conventional form.

\setcounter{equation}{0}
\section{New variables}
\subsection{Standard realization of conformal symmetry}
Let us start with the $so(1,2)$ algebra
\be\label{sl2}
\im \left[ H, D\right]=-H, \quad \im \left[ K, D\right]=K,\quad \im \left[ H,
K\right]=-2 D
\ee
The conformal symmetry can be realized on the
$SO(1,2)$ group element $g$
\be\label{g}
g=e^{\im t H} e^{\im z K}
e^{i u D} \ee by the left shifts \be\label{g0} g_0 \cdot g=
g',\qquad g_0=e^{\im a H} e^{\im c K} e^{i b D}
\ee
as
\be\label{conf2}
\delta t = a +b t +c t^2, \quad \delta u = b+ 2 c
t, \quad \delta z = c -\left( b + 2 c t \right) z.
\ee One should
stress that the transformation laws of $\dot u$
\be\label{dotu}
\delta \dot{u} = 2 c -\left( b+ 2 c t \right) \dot{u}
\ee
coincide
with those for $ 2 z$. These are the conventional transformation
properties of the fields under the one-dimensional conformal group
\cite{FF}.

One may extend the $so(1,2)$ algebra \p{sl2} by the additional
generator $U$, which commutes with the set $(H,D,K)$. The extended
group element ${\tilde g}$ will contain the additional factor
\be\label{tg}
{\tilde g}=e^{\im t H} e^{\im z K} e^{i u D}e^{i\theta U} ,
\ee
and the new bosonic field $\theta(t)$  will
have trivial transformation properties under the $SO(1,2)\times
U(1)$ group
\be\label{u1} \delta \theta = \gamma,
\ee
where $\gamma$ is the parameter of a $U(1)$ transformation.

All $SO(1,2)$ invariant objects can be constructed with the help
of Cartan forms \be\label{Cf1} {\tilde g}^{-1} d {\tilde g} = \im
\omega_H H+\im \omega_K K+\im \omega_D D +\im \omega_U U , \ee
where for our parametrization of $\tilde g$ \p{tg} the forms read
\be\label{Cf2} \omega_H= e^{-u} dt, \quad \omega_K = e^u\left( dz
+z^2 dt\right), \quad \omega_D=du - 2 z dt, \quad \omega_U=
d\theta. \ee

It is a matter of calculation to relate $(t, u, z)$ with the
coordinates $ (\tau, r, \phi)$ \be\label{rel1} t= \tau
+\frac{1}{r}, \quad u = - \log r - \phi, \quad z =\frac{r}{2} .
\ee In these variables the metric \p{metric1} acquires the form
\bea\label{metric3} ds^2& = &f_1 \left[ -4 dt \left( dz+ z^2
dt\right)+d\theta^2 \right]+
f_2\left[ d u -2 z dt \right]^2   \nn \\
&=& f_1 \left[ -4 \omega_P \; \omega_K +\omega_U^2 \right]+
f_2\omega_D^2 . \eea The full symmetry group leaving the metric
\p{metric3} invariant, includes also one additional isometry
\be\label{iso} \delta u= a_0, \ee which commutes with $SO(1,2)$
transformations \p{conf1}. It is worth to note that just this
additional isometry almost completely  fixed (up to the functions
$f_1,f_2$) the metric \p{metric3}.

Let us remind that the standard conformal mechanics is defined by
the action \cite{FF, leva} \be\label{stac} S_{stand} = \int dt
\left[ \alpha \omega_H +\beta \omega_K +\gamma \omega_D\right],
\ee where $(\alpha, \beta, \gamma)$ are arbitrary constant
parameters. In addition, the standard description is supplied by
the Inverse Higgs  condition \cite{IH} \be\label{ih1}
\omega_D=0\qquad \Rightarrow \qquad z=\frac{1}{2} {\dot u}, \ee
which expressed the additional Goldstone boson $z(t)$ in terms of
the dilaton $u(t)$. This is achieved without any breaking of
conformal symmetry due to the transformation properties \p{dotu}.
In contrast, in the action \p{metric3} the boson $z(t)$ itself is
present. Of course, one may additionally impose the condition
\p{ih1}; however, after this we will end up with three
two-dimensional systems. Thus, in what follows, we will avoid
imposing any additional constraints. Note, that even in this case
one may introduce, instead of the field $z(t)$, another one
defined as \be\label{newz} {\tilde z} = e^{u}\left( z -
\frac{1}{2}\,{\dot u}\right) \quad \Rightarrow \quad \delta
{\tilde z} =0. \ee Clearly, ${\tilde z}$ transforms as an ordinary
scalar field, like $\theta$, under the conformal $SO(1,2)$ group.

Another way to construct the conformal mechanics is to use the
following action
\be\label{altCM1}
S=-\gamma \int \sqrt{ \omega_H
\, \omega_K} = -\gamma \int dt \sqrt{\dot{z}+z^2}.
\ee
The equation of motion for a single variable $z$ reads
\be\label{altCM2}
\ddot{z} +6 z \dot{z} +4 z^3 =0.
\ee
The variable $z(t)$ is not very convenient: it has the unusual
dimension $(cm^{-1})$ and is shifted by a constant under the
conformal boost \p{conf2}. This is the property of the Goldstone
field for the conformal boost. Moreover, in higher dimensions this
variable is a vector with respect to the Lorentz group. All this
yields the motivation for introducing a new variable $x(t)$ as
\be\label{altCM3}
z(t) = \frac{d}{dt} \log x .
\ee
{}For this
variable the equation of motion \p{altCM2} will be of the third
order in time derivatives
\be\label{altCM4}
x \dddot{x}+ 3 \dot{x} \ddot{x} =0 .
\ee
Nevertheless, one may rewrite the equation
\p{altCM4} as
\be\label{altCM5}
\frac{d}{dt} \left( x^3 \ddot{x} \right) =0 \qquad \Rightarrow \qquad x^3 \ddot{x} = g^2 = const.
\ee
Keeping in mind that under the conformal group $x$ transforms as
\be\label{altCM6}
\delta x = \frac{1}{2} \left( b+ 2 c t\right) x,
\ee
one may check  that the combination $x^3 \ddot{x}$ is a
scalar under the conformal group and, therefore, the presence of
the constant $g^2$ in \p{altCM5} preserves the conformal symmetry.
Clearly, the equation \p{altCM5} is just an equation of motion of
the standard conformal mechanics \cite{FF}.

Another way to get the same result is to impose the Inverse Higgs
phenomenon condition \p{ih1} and then to represent the equation
\p{altCM2} as \be\label{zz}
 \frac{d}{dt} \left[ e^{\frac{3}{2}u} \frac{d^2}{dt^2}\left( e^{\frac{1}{2}u}\right)\right]=0 \qquad
 \Rightarrow \qquad \frac{d^2}{dt^2}\left( e^{\frac{1}{2}u}\right) =g^2 e^{-\frac{3}{2}u}.
\ee Clearly, this is the same equation as \p{altCM5} one, upon the
identification of $x$ with $e^{\frac{1}{2} u}$.

Thus, the action \p{altCM1} provides an alternative description of
conformal mechanics, when the coupling constant $g$ appears as an
integration constant.

\subsection{Particle action in new coordinates}
In newly defined coordinates \p{rel1}  the action \p{sAG} acquires
the form\footnote{The action, similar to \p{aSK},  has been
recently considered in \cite{Fed}.} \be\label{aSK} S=-m \int dt
\sqrt{ f_1 \left[ 4 \left( {\dot z}+z^2\right)
-{\dot\theta}{}^2\right] -f_2 \left( \dot u  - 2z\right)^2}. \ee
The generators of conformal transformations read \bea\label{genSK}
&& H=-\frac{ f_1\left( p_u^2+m^2 f_2\right)+f_2\left( p_\theta^2-2 z p_u p_z +z^2 p_z^2\right)}{f_2 p_z}, \nn \\
&& D=-p_u+z p_z +t H,\qquad K= -p_z +2 t D -t^2 H. \eea They form
the same $so(1,2)$ algebra \p{sl2AG} with respect to standard
Poisson brackets \be\label{pbSK} \left\{u ,p_u\right\}=1,\quad
\left\{z ,p_z\right\}=1,\quad \left\{\theta ,p_\theta\right\}=1.
\ee The Casimir element of the $so(1,2)$ algebra is simplified to
be \be\label{Casimir} J=2\left( H\; K- D^2\right) =2\left[ \left(
\frac{f_1}{f_2}-1\right) p_u^2 +p_\theta^2+m^2 f_1\right]. \ee
Now, following  \cite{AandAll}, we introduce the new canonical
variable $R(t)$ and its momentum $P_R$ as \be\label{R} R=\sqrt{2
K}, \qquad P_R =-\frac{\sqrt{2}D}{\sqrt{K}}\quad \rightarrow \quad
\left\{R,P_R\right\}=1. \ee With the help of these variables one
may rewrite our Hamiltonian \p{genSK} as \be\label{myH}
H=\frac{P_R^2}{2}+\frac{J}{R^2}. \ee This is the standard form of
any conformal invariant mechanics \cite{AandAll}. Thus, the task
of constructing a $N=2$ supersymmetric extension of our model is
reduced to a more simple problem -  the construction of the
supersymmetric extension of the system with the Hamiltonian
\p{myH}.

\setcounter{equation}{0}
\section{Supersymmetric extensions}
\subsection{Supersymmetrization of the ``radial'' variable}
If we will treat our Casimir operator $J$ \p{Casimir} as a
coupling constant, which does not participate in the
supersymmetrization procedure, then the Hamiltonian $H$ \p{myH} is
just the Hamiltonian of standard conformal mechanics \cite{FF}. In
this case $N=2$ superconformal extension of our system has the
same form as $N=2$ superconformal mechanics  \cite{BP}
\bea\label{n2Galaj}
&& {\cal Q}= \left( P_R+ \frac{\im \sqrt{2 J}}{R}\right) \psi, \quad \overline{\cal Q}= \left( P_R - \frac{\im \sqrt{2 J}}{R}\right) \bpsi,\qquad {\cal S}=-R \psi,\quad \overline{\cal S}=-R\bpsi,\nn\\
&&{\cal H}=\frac{P_R^2}{2}+\frac{J}{R^2}-\frac{\sqrt{2 J}}{R^2}
\psi\bpsi,\quad {\cal D}=-\frac{1}{2}R P_R ,\quad {\cal
K}=\frac{1}{2}R^2,\quad {\cal U}=\sqrt{2 J}+ \psi\bpsi. \eea Here,
we introduced the fermionic variables $(\psi,\bpsi)$ obeying the
following brackets \be \left\{ \psi, \bpsi \right\} = \im . \ee
One may easily check that all commutators of $N=2$ superconformal
algebra are satisfied\footnote{We write down only non-vanishing
brackets}: \bea\label{N2sca}
&&\left\{\cH, \cD \right\}=\cH, \quad \left\{\cK, \cD \right\}=-\cK, \quad \left\{\cH, \cK \right\}=2 \cD , \nn \\
&& \left\{ \left( \begin{array}{c} \cQ \\\overline\cQ \end{array}\right), \cD\right\}=\frac{1}{2}\left( \begin{array}{c} \cQ \\\overline\cQ \end{array}\right),\quad
\left\{ \left( \begin{array}{c} \cS \\\overline\cS \end{array}\right), \cD\right\}=-\frac{1}{2}\left( \begin{array}{c} \cS \\\overline\cS \end{array}\right), \nn\\
&& \left\{ \left( \begin{array}{c} \cQ \\\overline\cQ \end{array}\right), \cK\right\}=\left( \begin{array}{c} \cS \\\overline\cS \end{array}\right),\quad
\left\{ \left( \begin{array}{c} \cS \\\overline\cS \end{array}\right), \cH\right\}=-\left( \begin{array}{c} \cQ \\\overline\cQ \end{array}\right), \nn\\
&& \left\{ \left( \begin{array}{c} \cQ \\\overline\cQ \end{array}\right), \cU\right\}=\im\left( \begin{array}{c} -\cQ \\\overline\cQ \end{array}\right),\quad
\left\{ \left( \begin{array}{c} \cS \\\overline\cS \end{array}\right), \cU\right\}=\im\left( \begin{array}{c} -\cS \\\overline\cS \end{array}\right), \nn\\
&&\left\{ \cQ, \overline\cQ \right\}=2 \im \cH,\quad \left\{ \cS, \overline\cS \right\}=2 \im \cK, \qquad
\left\{ \cQ, \overline\cS \right\}=2 \im \cD +\cU,\quad \left\{ \overline\cQ, \cS \right\}=2 \im \cD -\cU.
\eea
This is the $N=2$ supersymmetrization constructed in \cite{AG}.

Clearly, this variant of supersymmetrization does not feel any
structure of the theory hidden in the Casimir $J$. This is just
the supersymmetrization of the ``radial'' variable $R$. It is
going to go in the same way for any $J$. The ``angular'' sector,
defined by the explicit structure of $J$, can contain as many
variables as we wish -- nothing will change. Moreover, the number
of supersymmetries is not restricted. Indeed, following
\cite{leva1,AN1} one may easily construct the $N=2 k$
superconformal invariant extension of this system for arbitrary
$k$ in a similar way. The corresponding superconformal group will
be the $SU(1,1|k)$ one.

\subsection{New variant of the supersymmetrization}

Another variant of the supersymmetrization we are going to
construct here, will include an additional $N=2$ supersymmetric
extension of the Casimir $J$. The present variant crucially
depends on the structure of the ``angular'' sector of the theory,
defined by $J$ \p{Casimir}. It is based on the same ideas as in
\cite{AandAll}. Generally, the construction goes as follows.

First of all, let us rewrite our bosonic Hamiltonian \p{myH} as
\be\label{myHa} H=\frac{P_R^2}{2}+\frac{m^2}{R^2}+\frac{\tilde
J}{R^2}, \qquad \tilde J = J-m^2 \ee Let us further suppose that
we were able to find a proper $N=2$ supersymmetric extension of
the ``angular Hamiltonian'' $J_{SUSY}={\tilde J}+\mbox{fermions}$.
This means that there are two supercharges $q,{\bar q}$ which
anti-commute to produce $J_{SUSY}$ (These supercharges will be
constructed in the next Section) \be\label{q} \left\{ q, {\bar
q}\right\}=2 \im J_{SUSY}=2 \im \left( {\tilde
J}+\mbox{fermions}\right), \qquad \left\{ q, q \right\}=\left\{
{\bar q}, {\bar q}\right\} =0. \ee We will also need the $U(1)$
generator $U_A$ for the supercharges $q, {\bar q}$ of the
``angular'' sector \be\label{addU} \left\{U_A, q\right\} = \im
\,q, \qquad \left\{U_A, {\bar q}\right\} = -\im \,{\bar q}. \ee
Clearly, $U_A$ is precisely the $R$-symmetry generator of $N=2$
supersymmetry. For the concrete system it has a standard
realization through the bilinear combinations of the fermions of
the ``angular'' sector (see the next Section). In the general
case, without going into the details of the particular model, one
may build this generator in terms of the supercharges $q,{\bar q}$
and the Hamiltonian $J_{SUSY}$ as \be\label{addU1} U_A=
\frac{q\,{\bar q}}{2\, J_{SUSY}}. \ee

Now, it is a matter of calculation to check that the following
generators form the same $N=2$ superconformal algebra \p{N2sca}
\bea\label{finN2SCA} && {\widetilde Q}=Q_0+\frac{q}{R}+\im
\frac{U_A \psi}{R},\quad \widetilde {\overline Q}={\overline
Q}_0+\frac{\bar q}{R}-\im \frac{U_A \bpsi}{R},\qquad
{\widetilde S}=-R \psi,\; \overline{\widetilde S}=-R \bpsi \nn \\
&& {\widetilde H}=H_0+\frac{\tilde J}{R^2}+\frac{1}{R^2}\left( \sqrt{2} \,m U_A-\im \psi {\bar q}-\im \bpsi q -U_A \psi\bpsi\right), \nn \\
&& {\widetilde U}=\sqrt{2}m +\psi\bpsi+U_A, \quad {\widetilde
K}=\frac{1}{2}R^2, \quad {\widetilde D}=-\frac{1}{2}R P_R, \eea
where \be\label{zeroN2} Q_0=\left(P_R+\im
\frac{\sqrt{2}m}{R}\right)\psi,\; {\overline Q}_0=\left(P_R-\im
\frac{\sqrt{2}m}{R}\right)\bpsi,\quad
H_0=\frac{1}{2}P_R^2+\frac{m^2}{R^2}-\frac{\sqrt{2}m
\psi\bpsi}{R^2}, \ee and, therefore, \be \left\{ Q_0,{\overline
Q}_0\right\}= 2 \im H_0. \ee Thus, the present construction
extends the arbitrary $N=2$ supersymmetric system with the
Hamiltonian $J_{SUSY}$ to the superconformal mechanics.

\subsection{Supersymmetric extension of the ``angular'' sector}
As it follows from the previous subsection, the $N=2$
superconformal extension of our system is reduced now to a much
simpler task -- the derivation of a $N=2$ supersymmetric extension
of the ``angular'' sector with the bosonic Hamiltonian
\be\label{angH} {\tilde J}= 2 \left( \frac{f_1}{f_2}-1\right)
p_u^2 +2 p_\theta^2 + m^2 \left( 2 f_1-1\right). \ee In order to
find a proper supersymmetrization of the system with this bosonic
Hamiltonian, let us firstly introduce a pair of complex fermionic
variables $\rho, \xi $ obeying the brackets \be \left\{ \rho,
\bar\rho \right\} = \im , \qquad \left\{ \xi, \bar\xi \right\} =
\im . \ee Now, it is rather easy to check that the supercharges
$q, {\bar q}$ \be\label{SCAn} q=2 p_\theta\, \rho +2 h_1\, p_u\,
\xi +\im \sqrt{2} m h_2\, \rho +\im h_3\, \xi \bar\xi \rho,\;
{\bar q}= 2p_\theta\,\bar\rho +2 h_1 \, p_u \,\bar\xi -\im
\sqrt{2} m \, h_2 \bar\rho - \im \, h_3 \xi \bar\xi \bar\rho, \ee
where the arbitrary, for the time being, functions $h_1, h_2$ and
$h_3$ depend on $\theta$ only, form the superalgebra \p{q} with
the Hamiltonian \be\label{HamA} J_{SUSY}=2 p_\theta^2 +2 h_1^2
p_u^2 +m^2 h_2^2+ 4\im  h_1' p_u \left( \rho
\bar\xi+\bar\rho\xi\right)-2 \sqrt{2} m \frac{h_1' h_2}{h_1}
\xi\bar\xi+ 2 \sqrt{2} m h_2' \rho\bar\rho -4 \left(
\frac{h_1'}{h_1}\right)' \rho\bar\rho\,\xi\bar\xi, \ee if the
function $h_3$ is defined as \be\label{sol} h_3= -2
\frac{h_1'}{h_1}. \ee Finally, identifying the bosonic part of
$J_{SUSY}$ \p{HamA} with $\tilde J$ \p{angH} we will fix all our
functions to be \be\label{sol2} h_1 = \sqrt{\frac{f_1}{f_2}-1},
\qquad h_2 = \sqrt{2 f_1 -1}. \ee It is worth to note that with
the newly introduced fermions $\rho, \xi$ the $U_A$ current
\p{addU} can be easily constructed as \be\label{UAf} U_A = \rho\,
\bar\rho + \xi\,\bar\xi. \ee Substituting the supercharges $q,
{\bar q}$ \p{SCAn} and the $U_A$ current \p{UAf} into the
supercharges and the Hamiltonian of the full system \p{finN2SCA},
we will get the desired $N=2$ supersymmetric extension of our
system. It has a proper number of fermions - two for each bosonic
variables, as it should be in $N=2$ supersymmetric mechanics. One
should stress that, with $m=0$, the bosonic Hamiltonian $\tilde J$
\p{angH} can be supersymmetrized, with the help of only two
fermions, within chiral superfields. However, for a generic value
of $m$, this is impossible and we have to invoke four fermions -
i.e. one complex fermions for each boson. The peculiarity of the
given system is encoded in the explicit expression for the
functions $h_1, h_2, h_3$ entering the supercharges \p{SCAn} and
the Hamiltonian \p{HamA}. In contrast, the supersymmetrization of
only the ``radial'' variable does not feel the fine structure of the
``angular'' sector and it has the same form \p{n2Galaj} for any
${\tilde J}$. That is the reason why the number of supersymmetries
can be chosen arbitrarily for the variant of the
supersymmetrization considered in \cite{AG}.

\setcounter{equation}{0}
\section{Conclusion and Discussion}
In the present paper, following \cite{AG}, we constructed a new
$N=2$ supersymmetric extension of a massive particle moving near
the horizon of the extreme Kerr black hole. The main motivation
for our paper was the unusual variant of supersymmetrization
proposed in \cite{AG}. Indeed, the supercharges and the
Hamiltonian constructed in \cite{AG} contain only two fermions,
while the number of physical bosons is three, in contrast with any
customary countings in one dimensional supersymmetric systems,
which all claim that the number of physical fermions cannot be
smaller than the number of physical bosons. We analyzed in details
this situation and explicitly demonstrated that the
supersymmetrization procedure used in \cite{AG} deals only with
the ``radial'' sector of the system. Such a supersymmetrization
leaves aside all peculiarities of the system encoded in its
``angular'' sector. We also chose a proper set of bosonic
variables with clear geometric meaning, and then we constructed
new $N=2$ supercharges and Hamiltonian which solved our task.

Clearly, the proposed approach can be applied to any system
possessing $SO(1,2)$ symmetry, including four additional
multi-parameters solutions of the vacuum Einstein equation we
found in  this paper. Clearly, the system recently presented in
\cite{AG1}, which suffers from the same problems as that in
\cite{AG}, can be easily re-formulated within our approach.

\section*{Acknowledgements}
We are indebted to Armen Nersessian for valuable discussions.

S.K. thanks the INFN-Laboratori Nazionali di Frascati, where this work was completed, for
warm hospitality.

This work was partly supported by RFBR grants  09-02-01209,
11-02-90445-Ukr, 11-02-01335, as well as by the ERC
Advanced Grant no. 226455, \textit{``Supersymmetry, Quantum Gravity and Gauge Fields''%
} (\textit{SUPERFIELDS}).

\end{document}